# Register-based Census in Thailand: a Case Study in Chachoengsao Province

Nuttirudee Charoenruk[1], Narongrid Asavaroungpipop[1], Pannee Pattanapradit[3], Kittiya Ku-kiattikun[2], and Chainarong Amornbunchornvej[2]*

[1] Department of Statistics, Chulalongkorn Business School, Chulalongkorn University, Thailand
[2] National Electronics and Computer Technology Center, Thailand
[3] National Statistical Office, Thailand
*Corresponding author: Chainarong.amo@nectec.or.th

**Abstract**
The use of registers has been increasingly popular in the field of population census because of its advantages over the traditional census. While the traditional census requires a large amount of fieldwork and data collection, the registered-based census can rely on pre-existing administrative data. As a result, the register-based census can save both time and budget. Thailand explored a use of the register-based census in 2020. In this paper, the authors layout the methodology in an aspect of data preparation and integration as well as analyze data quality of the register-based census compared with the traditional census in Chachoengsao province, Thailand. In addition, we compared conceptual frameworks that are commonly used for a register-based census in several countries and the number of databases (a recent single database VS multiple databases) used to construct the register-based census. We found that using a conceptual framework that counts the number of populations based on the main census variables on a single recent database is better than using a framework that counts population who appears on many registers in term of overcoverage and data distribution regarding to sex. This provides the evidence of using one recent and complete database is sufficient for conducting the register-based census. The authors end up with recommendations for conducing the register-based census.

**Acknowledgement**
We would like to thank the Thai National Statistical Office and National Electronics and Computer Technology Center who provided the authors data and facility to conduct the research.

## 1. Introduction

Population census has been conducted for many years with a goal of enumerating number of population and examining socio-demographic structure (Dygaszewicz, 2020). In Thailand, traditional census, by mainly using a face-to-face interview, has been conducted every 10 years from 1960 to 2010 (Grey et al., 2009). Data from the traditional census has been used to design country development plan and public policies, to estimate the number of populations, and to be used as a sampling frame for other surveys (Grey et al., 2009). In the past two decades, traditional census has encountered several problems. Respondents were not willing to participate in the traditional census (Dygaszewicz, 2020; Gisser, 2020). This leads to a decrease in response rate and low data quality (Grey et al., 2009). Additionally, it is costly to conduct a traditional census (Grey et al., 2009; Beltadze, 2020; Gisser, 2020). As such, instead of a traditional census, several countries have conducted register-based census in which administrative data from registers have been used.

To transmit from traditional census to register-based census, some countries are currently doing a combined census using both the administrative data and the traditional census. For example, in 2021,

Poland conducted the census by using a mixed model which incorporates administrative data and data collected directly from respondents (Statistics Poland 2021, Dygaszewicz 2020). Data from registers and data from respondents were compared to assess the data quality from registers. A list for 2021 Poland census is built based on list of persons with assigned address and list of the address and housing.

Currently, several countries especially in European countries altered the way to collect the census data from the traditional census to the register-based census (Dygaszewicz, 2020; Grey et al., 2009). For example, in 2011, Austria implemented a completely register-based census in which registers consist of a population census, a census of workplaces, building census, and housing census (Lenk, 2008; Gisser, 2020). In the same year, Sweden was conducted a register-based census using business register, population register, real property register, and dwellings register (Axelson et al., 2021).

Five conceptual frameworks that many countries have used to develop administrative data from registers are as follows:

*Framework 1.* Count population who can identify their current address. If a person has been identified on the current address in the nation, he/she will be included on the register-based census.

*Framework 2.* Count population who has Personal Identification number (PID) on registers. If a person has a PID on every register, he/she will be included on the register-based census.

*Framework 3.* Count population who has data on main census variables shown on registers. If a person has data on every main census variable (such as PID, full name, sex) on registers, he/she will be included on the register-based census. Countries must decide on census variables. In this study, we have used PID, First name, Last name, year of birth, prefix of name, and sex as the main census variables.

*Framework 4.* Count population who has at least some data on main census variables shown on register. If a person has at least some data on census variables on registers, he/she will be included on the register-based census. In this study, if a person has at least data for PID, First name, Last name, year of birth, prefix of name, or sex, he/she will be included on the register-based census.

*Framework 5.* Count population who appears on many registers. If a person appears on many registers, he/she will be included on the register-based census. In this study, we specified the number of registers to be at least 2.

Note. Some countries have been used more than one framework mentioned above.

In Thailand, due to the COVID-19 pandemic in 2020, National Statistic Organization (NSO) collected the traditional population census data only from 4 districts namely Panat Nikom, Chonburi; Sriracha, Chonburi; Rayong; and Chachoengsao. This study will focus only on Chachoengsao province which consist of 419,492 people in the population (214,044 females and 205,448 males) in the register-based census. The pyramid of population regarding age and sex shows in Figure 1. Note that, due to the COVID-19 outbreak, only partial areas of Chachoengsao were completely surveyed in the 2019 traditional census. Hence, we selected only areas that had been surveyed in traditional for register-based census. This made the total of population in the census areas was 182,614 populations.

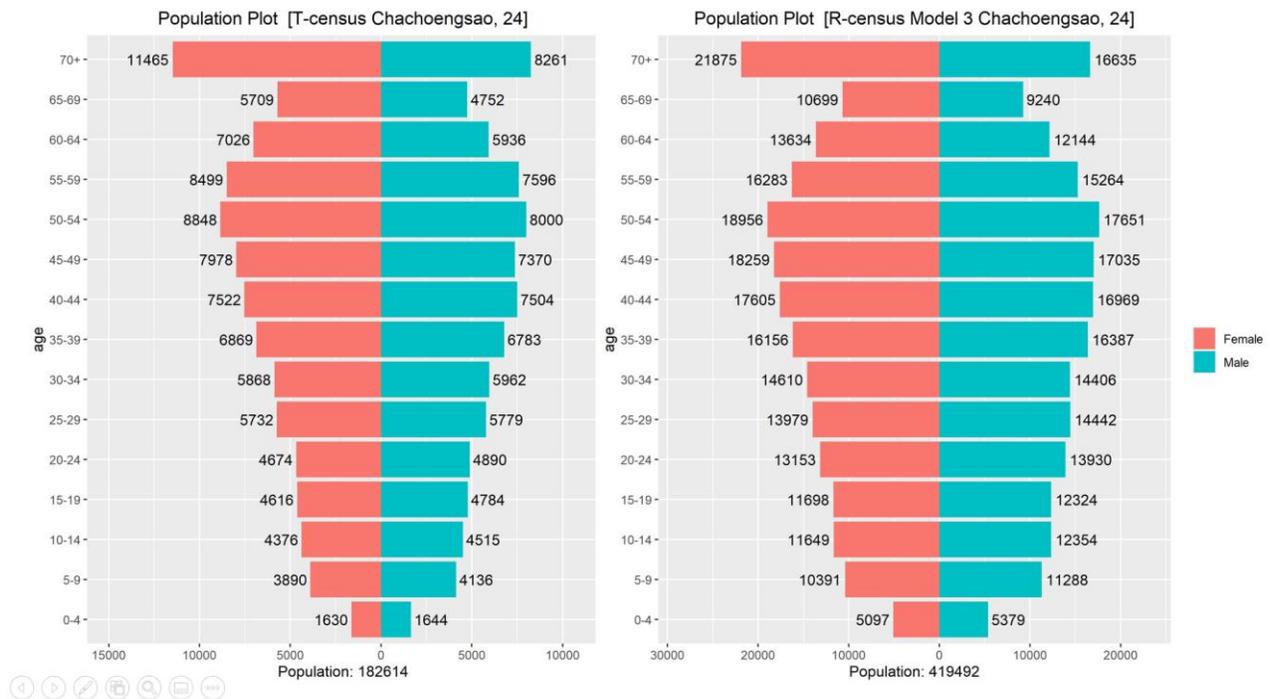

**Figure 1.** The pyramid of populations in Chachoengsao province. (Left) the pyramid from the traditional census. (Right) the pyramid from the register-based census.

This paper layouts methods for data preparation and integration for Register-based census. In addition, the authors evaluate the data quality of register-based census developed under the 5 conceptual frameworks and the number of databases (a recent database VS multiple databases) by comparing data from respondents in Chachoengsao with administrative data from registers. Four measures of data quality include coverage rate, overcoverage rate, Chi-square histogram distance of sex and age distribution between register-based census and traditional census have been used. In our conclusion and discussion, we address methodological and practical issues found from this study along with suggestion.

## 2. Methods
### 2.1 Method for data preparation and integration
**Data pipeline for R-census**

Before analyzing register databases, the first step is to standardize all available databases that came from various sources and have different formats to be the same format without any unusable values. Then, all databases are integrated together to be a single database that is ready for analyzing in the register-based census. To accomplish this goal, the raw databases have to be processed into the data pipeline.

A data pipeline is a process that takes raw databases as inputs and transforms them into output database(s) that are ready to be used for a specific purpose (Flynn et al, 2012). For the register-based census, the data pipeline consists of three important steps: 1) data preparation and cleansing, 2) de-identification for anonymizing sensitive data, and 3) data integration. These three steps are the main parts of preparing data for register-based census in Austria (Lenk, 2008), Netherlands (Bakker et al., 2014), and Poland (Dygaszewicz, 2020).

## 2.1.1 Data preparation and cleansing
### i. Data preparation

Given raw databases from various sources, the data preparation step has a main purpose to prepare these databases that might have different formats into a manageable format.

The first step is to make all databases have the correct format. For example, if a database has non-roman characters, then the step that is required here is to convert the database into UTF-8 encoding. For a non-binary files, they have to be converted into either text files (e.g. the Comma-separated values (csv) file format) or database file format. Some files also contain special characters that are in the wrong places and need to be removed before proceeding the next steps (e.g. newline character \n within a name column).

The second step is to make all databases have the correct records and columns. Some files might contain several pieces of information within a single column (e.g. title, name, and surname is in a single column). Hence, this single column must be split into multiple columns. Another common case is a file that has a record as a single string and has a "fixed-length" bytes for specific information (e.g. the first 14 bytes are reserved for ID and the next 20 bytes are for a name). This kind of file must be converted into a file that has multiple columns per record. Each column contains only a single piece of information (e.g. ID, name, surname). After all databases are ready, they will be processed into the data cleansing step.

### ii Data cleansing

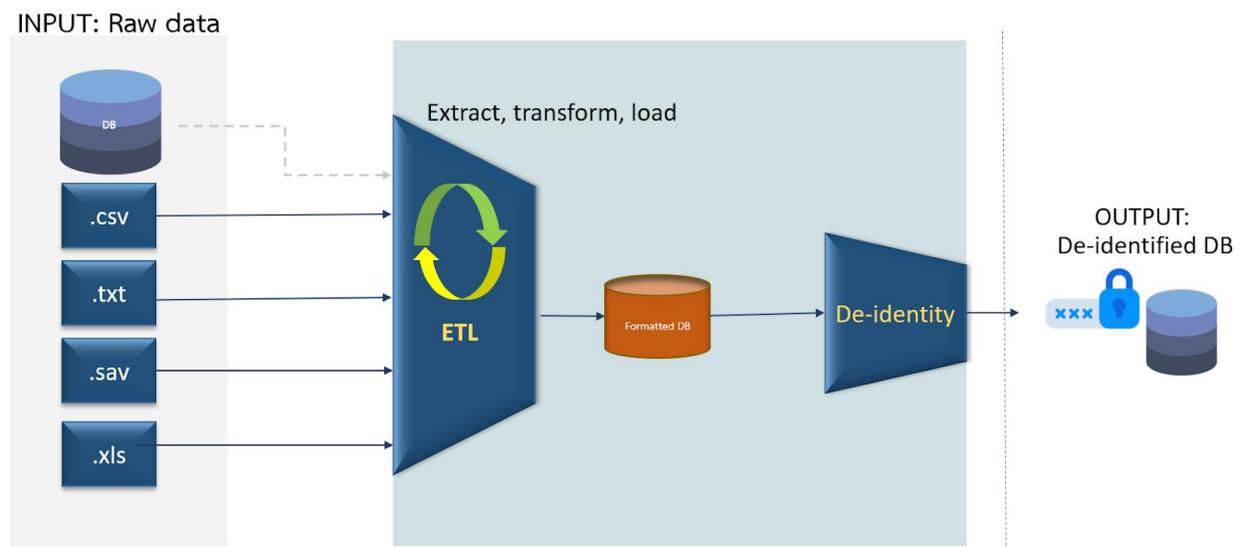

**Figure 2**. The steps of data preparation and cleansing.

In this step, the main goal is to make all values that share similar meaning have similar formats, and to remove all redundancy and unusable values within the databases so that these databases can be integrated into a single one without any problem.

An extract, transform, load (ETL) tool is the crucial one in this step. Briefly, the ETL tool is used to load data from different sources and transform them into the same format. For instance, in Figure 2, the databases are in different file formats. The ETL tool extracts these files into the process for transforming them to be formatted databases (e.g. .parquet or .csv files) and performing any actions to remove redundancy and unusable values before loading them into the next process, which is the de-identifying process.

For standardizing values with a similar meaning, the first step is to interpret all data dictionaries to find values that share similar meaning. For example, in a sex column, each database might represent as 'M' for male and 'F' for female, while some other databases might have '1' for male and '2' for female. If the 'M' for male and 'F' for female have been chosen as a standard, then the ETL tool must transform all sex columns to have only 'M' and 'F' values.

For redundancy, for each single database, if there are multiple rows with the similar primary key (e.g. citizen ID), then there are many actions that can be taken. If the redundant rows share similar name, surname, and other important values, we might keep the most updated one and remove others. If we cannot identify which row is more reliable than others, then we have to remove all of them. Otherwise, these redundancy rows can cause the overcoverage issue when joining a database to others since some people from redundant rows might not exist.

For unusable values, some databases were collected by typing, which can cause some typos issues. Hence, all columns might be transforming so that only valid values exist in the columns. For example, in the sex column, suppose only 'M' and 'F' are valid values, the ETL tools will remove anything else and replace those invalid values with N/A.

In this work, the Pentaho Data Integration (PDI) was deployed as the main ETL tool. The PDI is an open source ETL tool with high efficiency in resource utilization compared against several open source ETL tools (Majchrzak, 2011).

### 2.1.2 De-identification

The next step after data preparation and cleansing is the de-identification step. De-identification is a process that has the main purpose to protect sensitive information (e.g. citizen ID, name, medical records) while maintaining uniqueness of values so that these encrypted values can be used as keys for joining databases. De-identification uses a Cryptographic Hash Function (CHF) that can encrypt information such that it is almost impossible to decode the encrypted information.

In some countries, as required by laws, sensitive information of citizens such as personal data must be anonymized before being used for the census. Hence, the de-identification step is a crucial step for register-based census in some EU countries, such as Austria (Lenk, 2008), Netherlands (Bakker et al, 2014), etc.

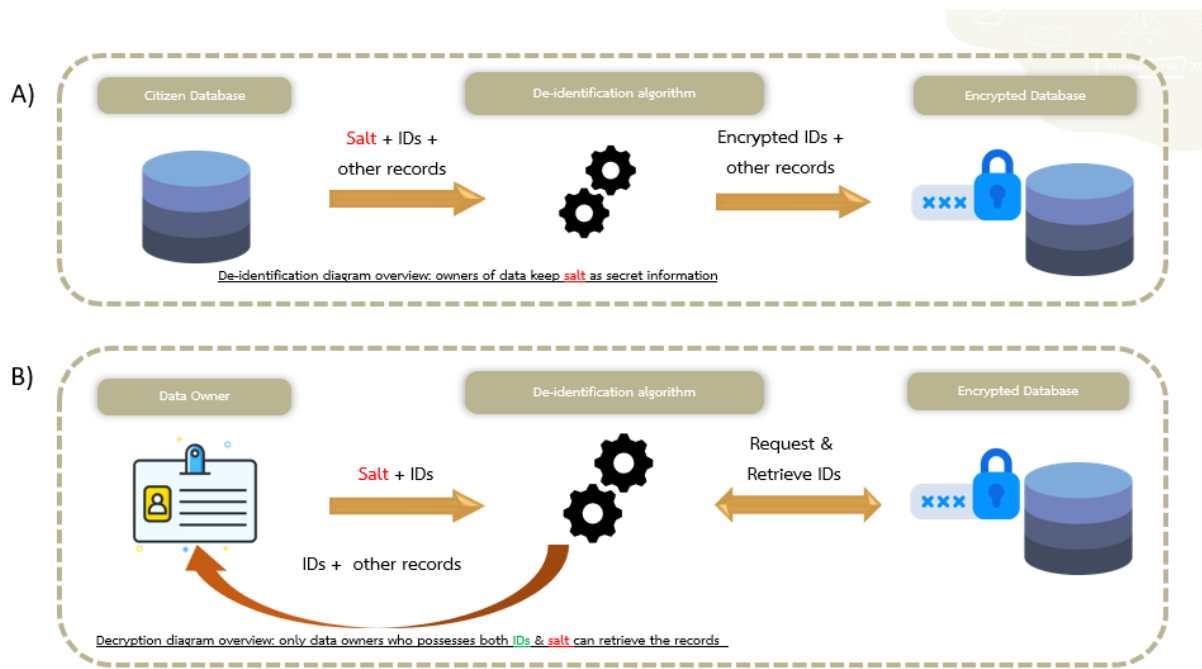

**Figure 3.** A) The steps of de-identification of a citizen database, which contains sensitive information. B) The steps of retrieving encrypted records back require a real ID from the data owner and the salt.

For the process of de-identification, there are two important steps: 1) encrypting information and 2) retrieving encrypted records with known keys. For the encrypting information step, in Figure 3 A), the input is a citizen database that contains sensitive information (citizen IDs, name, surname, etc.). Suppose only citizen IDs will be encrypted, in the first step, the process combines citizen IDs with the salt, which is a secret key. Then, the IDs with salt are encrypted by some de-identification algorithm. Finally, all original IDs are replaced with the encrypted IDs in the output database.

For retrieving encrypted records, in Figure 3 B), the database owner must have two things to retrieve the encrypted records: the salt (secret key) and the original ID before encryption. This typically can be done when a citizen who is the owner of ID and the database owner who possesses the salt both agree to retrieve the record. After having both salt and original ID, the next step is to combine salt and original ID with the same de-identification algorithm to get the encrypted ID. Then, by using the encrypted ID to search in the encrypted database, the target record can be retrieved.

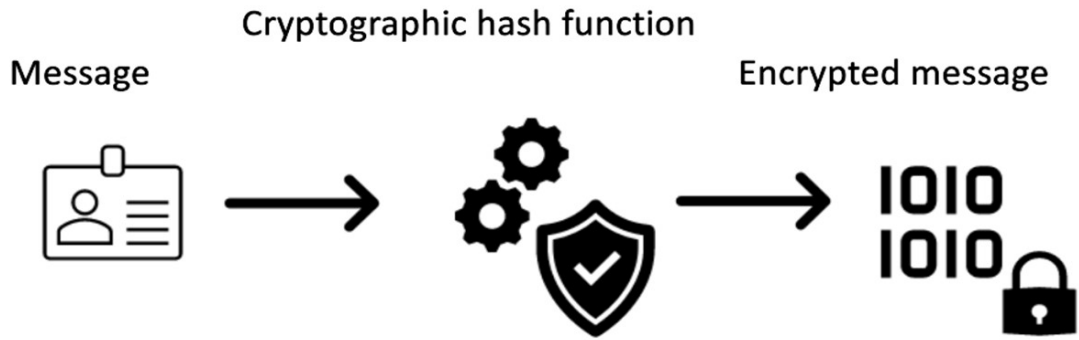

**Figure 4.** The process of encryption using a Cryptographic Hash Function (CHF)

In this work, we deploy the "SHA-256" method, which is a Cryptographic Hash Function (CHF) to perform de-identification. The CHF were invented and mainly used in the field of Cryptography to encrypt the original message to be encrypted one that is unable to decode back to the original value (Figure 4). The SHA-256 encrypts information using Secure Hash Algorithm 2 (SHA-2) with 256 bits of hash values. The SHA-2 was invented by the United States National Security Agency (NSA) in 2001 with the US patent 6829355 (Lilly, 2004).

For the SHA-256, it has 256 bits of hash values, which implies that there are $2^{256}$ possible values of encrypted values. No matter how long the original message is, after encryption, the encrypted message always has 256 bits of length. Compared to the popular encryption method, MD5 (Ciampa, 2009), which only has 128 bits of hash values, the SHA-256 is more secure in the term of decoding.

### 2.1.3 Data integration

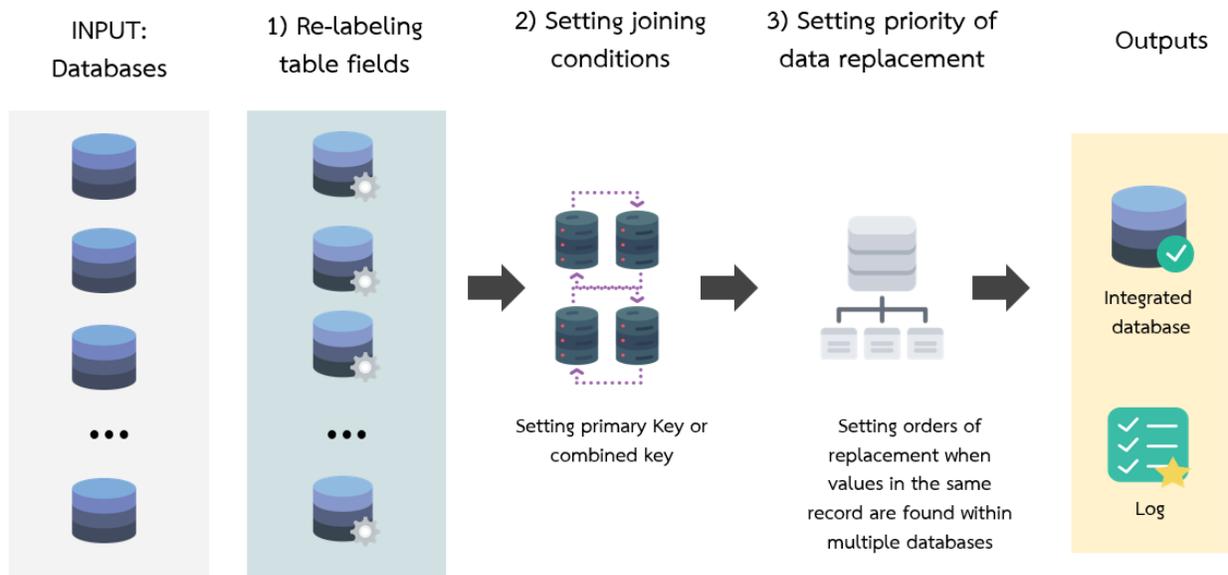

**Figure 5.** data integration steps

1. Data integration process

After all databases have been cleaned and de-identified, the next step before the analysis of register-based census is to integrate all databases together. In Figure 5, the process of data integration consists of three important steps of setting: 1) re-labeling table fields, 2) setting joining conditions, and 3) setting priority of data replacement.

In the step of re-labeling fields, since databases from different sources typically have the different field names that share the same meaning, we need to normalize them to be in the same standard. For example, some databases have the field 'sex' to represent the biological sex of citizen while other have 'gender' to represent the same thing. In this step, the main focus is to make the match between fields that represent the same meaning across databases. This can be done by analyzing the data dictionary of databases.

For the setting of joining conditions, the main goal is to define the primary key (or combined key) for joining all databases together. In the register-based census literature, the common field that is used as a primary key is the field of citizen ID because the IDs are unique among citizens, and they are typically used in registers as a primary key and as identifiers when retrieving citizen records. However, in some countries that have no citizen IDs, other fields might be used as keys.

Lastly, setting priority of data replacement among databases is the important step in data integration. When the same person has multiple records across databases, the issue of which values from which database should be used in the integrated database can be solved by using the priority list of data replacement. For example, suppose person A has a sex value as 'male' in four databases and 'female' in one database. The priority list is deployed to decide whether A is male or female based on the setting.

After all settings have been set, the databases will be integrated together to be a single database with some additional log data and ready to be used in the census analysis.

In the next subsection, the details of the three steps above are explained.

2. Primary Key

The primary key or combined key is used for joining tables from multiple databases. The following properties are necessary for considering any field(s) to be the primary key (Dygaszewicz, 2020).

- There is only one unique value for each person or object (e.g. institute, building, company) in the field(s).
- No two objects share the same value.
- The same value appears in all databases to refer to the same object.
- The value remains attached to the object forever; even if the object is obsoleted (e.g. a person is dead), there are no reusable values in the databases.
- The value can be validated with any digital check or reference register whether the value is correct. In the aspect of citizen IDs, the validation goal is to check whether the owner of the ID exists as a citizen in the country.

The citizen ID typically passes all criteria above, hence, it can be used as a primary key to join databases. In the case when citizen IDs are not available, the typical combined key that is used as a primary key is the combination of name, surname, date of birth, sex, and address fields.

The mandatory property beside above properties for joining databases is that all records must have values, otherwise, we have to remove records that have no value for a primary key. In the case of having a primary key as a citizen ID field, we have to remove all records in any databases that have no citizen IDs. In the case of having the combination of name, surname, date of birth, sex, and address fields as a parimary key, if any values in these fields is not available in any records, these record must also be removed. Choosing the fields to be a part of combined key can cause a problem if some fields (e.g. sex, date of birth) have too much empty values because there are many records that must be removed. Hence, if the citizen IDs or any identifier exists for all members in the population, then it should be used as a primary key since it can satisfy all properties above and, theoretically, there are no records that are needed to be removed.

For the foreigners or temporary visitors that are considered to be members of census population, it is challenging to count them since they might not have a citizen ID in some country (e.g. Thailand), and the universal ID might not exist to identify them. The field of visitor passport IDs might be used as a primary key in this case.

3. Joining conditions

After having primary key(s), the next step is to set the joining conditions of databases. In the case of one primary key, all databases can be simply joined with the key. However, if there are more than one primary key, the order of joining must be considered. For example, suppose there are two primary keys: the citizen-ID field, and the combined key of name, surname, and address, then we need a global primary key for integrating these primary keys together. The reason is that some records might have no value in the first primary key but have values in other primary keys. Hence, to satisfy the condition that a primary key must have value in all records, a global primary key might be created for each record as a new one.

If the primary key that has the highest priority is the citizen-ID field and the second priority key is the combined key, then there are three steps as follows. First, all databases must be joined with the citizen IDs and all records that cannot be joined as the first step (e.g. records without citizen IDs) must be moved to the next step. Then, new values in the global primary key are created for these records linking with the citizen IDs: each unique citizen ID has a unique global primary key. For the second step, the records that cannot be joined in the first step are joined together using the combined key. Then, a unique value of the global primary key is created and linked with each unique value of the combined key (i.e. a unique combination of name, surname, and address). Lastly, in the third step, all records in the first and second steps are merged together to be a single table.

In this work, we deploy the citizen ID field as a primary key.

4. Priority order of data replacement

When there are more than one records of the same person that have values in a specific field (e.g. sex), we have to decide that which value is the valid one. According to the work in Dygaszewicz (2020), the following properties can be considered when selecting value as a representative value in the integrated database.
- Timeline: the recent data comes first in the data replacement. The older data must be dropped.
- Relevance of databases: choosing values from reliable sources (Lenk, 2008). Some registers are the main register of a country that are maintaining citizen information (e.g. the register from the ministry of interior), while other databases are from less important ministries that do not have a direct duty to maintain valid citizen information. For example, if the record has a conflict of sex values between the database of the ministry of interior and databases from other ministries, then we choose the value from the ministry of interior register as the valid value.

- Coverage of information: choosing databases that are rich in the number of records and fields. In this method, large databases are considered to be more reliable than the smaller one.
- Quality of databases: choosing values from high quality databases first. The quality of databases can be measured in several ways. Below are some examples.
    - The completeness of fields: counting usable values in each field compared to the total number of records.
    - Timeline: how recent the data is.
    - Degree of integration: how many records have valid IDs w.r.t. some reference register(s).
    - Utility: how many fields are useful for the census analysis.

In this work, we deploy the timeline as the data replacement method since there is no other information regarding relevance of databases and all registers have almost the same size and quality.

**2.2 Method for examining data quality in register-based census**

To evaluate the 5 conceptual frameworks used to develop register-based census, the authors examine the data quality of register-based census by comparing traditional census collected from respondents and data from register-based census using 5 frameworks based on 4 measures as follows:

1. Coverage rate is the proportion of people who are in traditional census are actually in registers. This can be calculated from

$$= \frac{\text{number of people who are in registrer-based census }(R) \text{ and traditonal census }(T)}{\text{number of people who are in traditional census }(T)} = \frac{T \cap R}{T}$$

2. Overcoverage rate is the proportion of people who should not be in registers; however, they are in. This can be calculated from

$$= \frac{\text{number of people who are in register-based census }(R) \text{ but not in traditonal census}(T')}{\text{number of people who are in register-based census }(R)} = \frac{T' \cap R}{R}$$

3. Chi-square histogram distance measured a similarity of sex distribution between register-based census and traditional census and 4. Chi-square histogram distance measured a similarity of age distribution between register-based census and traditional census

Chi-square histogram distance is a measure used to examine the distance of data distribution between two sources, i.e. register-based census and traditional census. The chi-square histogram distance can be calculated by taking a square root of

$$\chi^2 = \frac{1}{2} \sum_{i=1}^{n} \frac{(p(i) - q(i))^2}{p(i) + q(i)}$$

where  $p(i)$ = proportion of population who are in a category $i$ of population in register-based census
$q(j)$ = proportion of population who are in a category $i$ of population in traditional census
$n$ = number of categories

An ideal register-based census should be indicated by having high coverage rate, low overcoverage rate, and small value of chi-square histogram distance for sex and age distribution.

# 3. Results

This paper compares the Register-based Census (R-census) and the Traditional census (T-census) by focusing on data coverage and the similarities in basic demographic characteristics including sex and age of Thai people in Chachoengsao province. To compare the use of the single recent database and the multiple databases to create R-census, two types of databases are used to be the registered based census which are the 2019 Basic Minimum Need (BMN) database (the 2019 R-census) and the 2017-2019 Basic Minimum Need (BMN) database (the 2017-2019 R-census).

The 2019 R-census and the 2017-2019 R-census are used to enumerate the numbers of members, i.e. Thai population, under 5 conceptual frameworks:

Framework 1. Count population who can identify their current address
Framework 2. Count population who has Personal Identification number (PID) on registers
Framework 3. Count population who has data on main census variables shown on registers
Framework 4. Count population who has at least some data on main census variables shown on register
Framework 5. Count population who appears on many registers (only for the 2017-2019 R-census)

Data obtained from each framework from the 2019 R-census and the 2017-2019 R-census were compared with data obtained from the 2019 population census database of the National Statistical Office (T-census) in Chachoengsao province, Thailand. The results of the comparison are presented in the following measures:

    1) Coverage rate
    2) Overcoverage rate
    3) Chi-square histogram distance measured a similarity of sex distribution (CHD-sex)
    4) Chi-square histogram distance measured a similarity of age distribution (CHD-age)

## 3.1 Results from analyzing only one recent registered database- the 2019 BMN database (the 2019 R-census)

Coverage and overcoverage between the 2019 R-census and data obtained from the traditional population census database of the National Statistical Office in 2019 (T-census) as well as the number of members in Chachoengsao province across framework 1-4 are reported in Table 1 and 2. As shown in Table 1, number of members across 4 frameworks in both coverage aspect (number of members in both the 2019 R-census and the T-census, R ∩ T) and overcoverage aspect (number of members in the 2019 R-census but not in the T-census, R ∩ T'.) are the same. This is because there is no missing data with respect to the main census variable including PID card number, first name, last name, year of birth, prefix, and sex in the 2019 BMN database.

Consequently, the coverage rate and overcoverage rate are identical across 4 frameworks. The coverage rate is 0. 689, reflecting the use of data from the 2019 R-census to represent 68.5% of the T-census data. The overcoverage rate is 0.656. The use of the 2019 R-census data to represent T-census data results in bias due to including data from the population in R-census but not in T-census, which accounts for 65.6% of the population in R-census.

Additionally, the number of Thai population as well as with respect to sex (Table 1) and age (Table 2) are the identical in framework 1-4. This reflects the completion in data collection on every main census variable. This is confirmed by interviewing with the data controller who manage the 2019 BMN database. The results show the flexibility in selecting any framework (Framework 1 to Framework 4) for creating the 2019 R-census database.

**Table 1.** Population size, coverage, overcoverage, and number of members under 4 frameworks in the 2019 R-census

| Census source | Framework | Population size | Coverage | | Overcoverage | | Sex | |
|---|---|---|---|---|---|---|---|---|
| | | | Number of member | Coverage rate | Number of member | Overcoverage rate | Female | Male |
| | | | T ∩ R | (T ∩ R)/T | (R ∩ T') | (R ∩ T') / R | | |
| 2019 R-Census | 1 | 365,394 | 125,729 | 0.689 | 239,665 | 0.656 | 186,601 | 178,793 |
| | 2 | 365,394 | 125,729 | 0.689 | 239,665 | 0.656 | 186,601 | 178,793 |
| | 3 | 365,394 | 125,729 | 0.689 | 239,665 | 0.656 | 186,601 | 178,793 |
| | 4 | 365,394 | 125,729 | 0.689 | 239,665 | 0.656 | 186,601 | 178,793 |
| T-Census | | 182,614 | | | | | 94,702 | 87,912 |

Note. Conceptual framework 5 will be used when the R-census is constructed from multiple registers. Since the single recent register has been used, the conceptual framework 5 cannot be used.

**Table 2.** Number of members across age under 4 frameworks in the 2019 R-census and the 2019 T-census

| Census source | | 2019 R-census | | | | T-census |
|---|---|---|---|---|---|---|
| Framework | | 1 | 2 | 3 | 4 | |
| Population size | | 365,394 | 365,394 | 365,394 | 365,394 | 182,614 |
| Age (year) | 0-4 | 2,830 | 2,830 | 2,830 | 2,830 | 3,274 |
| | 5-9 | 16,642 | 16,642 | 16,642 | 16,642 | 8,026 |
| | 10-14 | 20,112 | 20,112 | 20,112 | 20,112 | 8,891 |
| | 15-19 | 21,166 | 21,166 | 21,166 | 21,166 | 9,400 |
| | 20-24 | 22,325 | 22,325 | 22,325 | 22,325 | 9,564 |
| | 25-29 | 24,021 | 24,021 | 24,021 | 24,021 | 11,511 |
| | 30-34 | 24,669 | 24,669 | 24,669 | 24,669 | 11,830 |
| | 35-39 | 26,947 | 26,947 | 26,947 | 26,947 | 13,652 |
| | 40-44 | 29,722 | 29,722 | 29,722 | 29,722 | 15,026 |
| | 45-49 | 30,453 | 30,453 | 30,453 | 30,453 | 15,348 |
| | 50-54 | 32,504 | 32,504 | 32,504 | 32,504 | 16,848 |
| | 55-59 | 30,027 | 30,027 | 30,027 | 30,027 | 16,095 |
| | 60-64 | 24,842 | 24,842 | 24,842 | 24,842 | 12,962 |
| | 65-69 | 19,535 | 19,535 | 19,535 | 19,535 | 10,461 |
| | >=70 | 39,599 | 39,599 | 39,599 | 39,599 | 19,726 |

Note. Conceptual framework 5 will be used when the R-census is constructed from multiple registers. Since the single recent register has been used, the conceptual framework 5 cannot be used.

The difference in the population proportion with respect to sex and age between the 2019 R-census and the 2019 T-census is examined by using Chi-Square Test for Homogeneity and Chi-Square Histogram distance (CHD) and the result was shown in Table 3 and 4 respectively. According to sex, result from the Chi-Square Test for Homogeneity shows a statistically significant difference in the population proportion between the 2019 R-census and the 2019 T-census (p-value<0.001), indicating that the proportion of the population with respect to sex from the 2019 R-census is biased. The bias was measured by Chi-Square Histogram distance, which had a value of $6.257 \times 10^{-5}$. Alternatively, considering the median, the difference between the proportions of male and female population from the 2019 R-census and the T-census was found to be 0.79%.

From table 4, the Chi-Square Test for Homogeneity reveals that the proportion of the population categorized by age between the 2019 R-census and the T-census was statistically significantly different at the 0.05 level of significance (p-value < 0.001), indicating that the proportion of the population categorized by age from the 2019 R-census is biased. The bias was measured by Chi-Square Histogram distance, which had a value of 0.00305. Alternatively, considering the median, the difference between the proportions of population categorized by age from the 2019 R-census and the T-census was found to be 0.30%.

**Table 3.** Chi-square Test for Homogeneity with respect to sex comparing between the 2019 R-census and the T-census and Chi-square histogram distance

| Framework | Sex | R-census 2019 frequency (percentage) | T-census frequency (percentage) | Absolute Percentage difference | Chi-Square Statistics* (df) | P-value | CHD (SEX) |
|---|---|---|---|---|---|---|---|
| 1-4 | Female | 186,601 (51.07%) | 94,702 (51.86%) | 0.79% | 30.439 (1) | < 0.001 | 6.257E-05 |
|  | Male | 178,793 (48.93%) | 87,912 (48.14%) | 0.79% |  |  |  |

* Chi-Square Test for Homogeneity

## 3.2 Results from analyzing multiple registered databases- the 2017-2019 BMN database (the 2017-2019 R-census)

Table 5 and 6 show the coverage and overcoverage between the 2017-2019 R-census and data obtained from the 2019 traditional population census database of the National Statistical Office (the T-census) as well as the number of members in Chachoengsao province across framework 1-5. Coverage rate and overcoverage rate were identical from conceptual framework 1 to conceptual framework 4, which was 0.721 and 0.686 respectively, and for conceptual framework 5, the coverage rate and overcoverage rate were 0.682 and 0.683 respectively. Since there is no item nonresponse on the main census variables in the 2017-2019 R-census, the number of member as well as coverage and overcoverage are identical among the conceptual framework 1-4. Therefore, results in this section will be discussed separately into two groups which are conceptual frameworks 1-4 and conceptual framework 5.
The coverage rate in the conceptual frameworks 1-4 and the conceptual framework 5 was 72.1% and 68.2% respectively. This reflects the use of data from the 2017-2019 R-census to represent the T-census data from the conceptual frameworks 1-4 was better than that of the conceptual framework 5. In addition, the overcoverage rate in the conceptual frameworks 1-4 and the conceptual framework 5 was

**Table 4.** Chi-square Test for Homogeneity with respect to age comparing between the 2019 R-census and the T-census and Chi-square histogram distance

| Framework | Age | R-census 2019 frequency (percentage) | T-census frequency (percentage) | Absolute Percentage difference | Chi-Square Statistics* (df) | P-value | CHD (AGE) |
|---|---|---|---|---|---|---|---|
| 1 - 4 | 0-4 | 2,830 (0.77%) | 3,274 (1.79%) | 1.02% | 1,629.30 (14) | < 0.001 | 0.003051 |
|  | 5-9 | 16,642 (4.55%) | 8,026 (4.40%) | 0.16% |  |  |  |
|  | 10-14 | 20,112 (5.50%) | 8,891 (4.87%) | 0.64% |  |  |  |
|  | 15-19 | 21,166 (5.79%) | 9,400 (5.15%) | 0.65% |  |  |  |
|  | 20-24 | 22,325 (6.11%) | 9,564 (5.24%) | 0.87% |  |  |  |
|  | 25-29 | 24,021 (6.57%) | 11,511 (6.30%) | 0.27% |  |  |  |
|  | 30-34 | 24,669 (6.75%) | 11,830 (6.48%) | 0.27% |  |  |  |
|  | 35-39 | 26,947 (7.37%) | 13,652 (7.48%) | 0.10% |  |  |  |
|  | 40-44 | 29,722 (8.13%) | 15,026 (8.23%) | 0.09% |  |  |  |
|  | 45-49 | 30,453 (8.33%) | 15,348 (8.40%) | 0.07% |  |  |  |
|  | 50-54 | 32,504 (8.90%) | 16,848 (9.23%) | 0.33% |  |  |  |
|  | 55-59 | 30,027 (8.22%) | 16,095 (8.81%) | 0.60% |  |  |  |
|  | 60-64 | 24,842 (6.80%) | 12,962 (7.10%) | 0.30% |  |  |  |
|  | 65-69 | 19,535 (5.35%) | 10,461 (5.73%) | 0.38% |  |  |  |
|  | >= 70 | 39,599 (10.84%) | 19,726 (10.80%) | 0.04% |  |  |  |

* Chi-Square Test for Homogeneity

68.6% and 68.3% respectively. This indicates that the use of data from R-census to represent T-census data leds to a bias due to overcounting the population in the 2017-2019 R-census that was not in the T-census; however, the overcoverage rate do not significantly differ between the conceptual frameworks 1-4 and the conceptual framework 5.

**Table 5.** Population size, coverage, overcoverage, and number of members under 4 frameworks in the 2017-2019 R-census

| Census source | Framework | Population size | Coverage | | Overcoverage | | Sex | |
|---|---|---|---|---|---|---|---|---|
| | | | Number of members | Coverage rate | Number of members | Overcoverage rate | Female | Male |
| | | | $T \cap R$ | $(T \cap R)/T$ | $R \cap T'$ | $(R \cap T')/R$ | | |
| 2017-2019 R-Census | 1 | 419,492 | 131,592 | 0.721 | 287,900 | 0.686 | 214,044 | 205,448 |
| | 2 | 419,492 | 131,592 | 0.721 | 287,900 | 0.686 | 214,044 | 205,448 |
| | 3 | 419,492 | 131,592 | 0.721 | 287,900 | 0.686 | 214,044 | 205,448 |
| | 4 | 419,492 | 131,592 | 0.721 | 287,900 | 0.686 | 214,044 | 205,448 |
| | 5 | 393,034 | 124,467 | 0.682 | 268,567 | 0.683 | 200,588 | 192,446 |
| T-Census | | 182,614 | | | | | 94,702 | 87,912 |

**Table 6.** Number of members across age under 4 frameworks in the 2017-2019 R-census

| Census source | | 2017-2019 R-Census | | | | | T-Census |
|---|---|---|---|---|---|---|---|
| Framework | | 1 | 2 | 3 | 4 | 5 | |
| Population size | | 419,492 | 419,492 | 419,492 | 419,492 | 393,034 | 182,614 |
| Age (year) | 0-4 | 10,476 | 10,476 | 10,476 | 10,476 | 8,402 | 3,274 |
| | 5-9 | 21,679 | 21,679 | 21,679 | 21,679 | 19,727 | 8,026 |
| | 10-14 | 24,003 | 24,003 | 24,003 | 24,003 | 22,206 | 8,891 |
| | 15-19 | 24,022 | 24,022 | 24,022 | 24,022 | 22,381 | 9,400 |
| | 20-24 | 27,083 | 27,083 | 27,083 | 27,083 | 25,028 | 9,564 |
| | 25-29 | 28,421 | 28,421 | 28,421 | 28,421 | 26,271 | 11,511 |
| | 30-34 | 29,016 | 29,016 | 29,016 | 29,016 | 26,887 | 11,830 |
| | 35-39 | 32,543 | 32,543 | 32,543 | 32,543 | 30,431 | 13,652 |
| | 40-44 | 34,574 | 34,574 | 34,574 | 34,574 | 32,696 | 15,026 |
| | 45-49 | 35,294 | 35,294 | 35,294 | 35,294 | 33,402 | 15,348 |
| | 50-54 | 36,607 | 36,607 | 36,607 | 36,607 | 34,879 | 16,848 |
| | 55-59 | 31,547 | 31,547 | 31,547 | 31,547 | 30,158 | 16,095 |
| | 60-64 | 25,778 | 25,778 | 25,778 | 25,778 | 24,719 | 12,962 |
| | 65-69 | 19,939 | 19,939 | 19,939 | 19,939 | 19,189 | 10,461 |
| | >=70 | 38,510 | 38,510 | 38,510 | 38,510 | 36,658 | 19,726 |

The difference in the population proportion between the 2017-2019 R-census and the T-census with respect to sex and age groups were tested by using the Chi-Square Test for Homogeneity and the Chi-Square Histogram distance (CHD) and the results are in Table 7 and 8. As shown in the result from the Chi-Square Test for Homogeneity in Table 7, the population proportions are significantly different between the 2017-2019 R-census and the T-census in all of 5 conceptual frameworks (p-value < 0.001). This indicates that the proportion of the population classified by sex from the 2017-2019 R-census is biased. The bias can be measured by the Chi-Square Histogram distance, which had a value of 6.97037E-05 for the frameworks 1 to 4 and a value of 6.78432E-05 for the framework 5. The median of the difference between the population proportion of the male and female of the R-census and T-census were 0.84% and 0.82% respectively.

**Table 7.** Chi-square Test for Homogeneity with respect to sex comparing between the 2019-2019 R-census and the T-census and Chi-square histogram distance

| Framework | Sex | R-census frequency (percentage) | T-census frequency (percentage) | Absolute Percentage difference | Chi-Square Statistics* (df) | P-value | CHD (SEX) |
|---|---|---|---|---|---|---|---|
| 1-4 | Female | 214,044 (51.02%) | 94,702 (51.86%) | 0.84% | 35.467 (1) | < 0.001 | 6.97037E-05 |
|  | Male | 205,448 (48.98%) | 87,912 (48.14%) | 0.84% |  |  |  |
| 5 | Female | 200,588 (51.04%) | 94,702 (51.86%) | 0.82% | 33.830 (1) | < 0.001 | 6.78432E-05 |
|  | Male | 192,446 (48.96%) | 87,912 (48.14%) | 0.82% |  |  |  |

* Chi-Square Test for Homogeneity

As shown in Table 8, the Chi-Square Test for Homogeneity indicates the population proportion classified by age differ significantly between the 2017-2019 R-census and T-census in all frameworks (p<0.001). This indicates that the population proportion classified by age from the 2017-2019 R-census is biased. The bias can be measured by the Chi-Square Histogram distance, which had a value of 0.004243 for the framework 1 to 4 and a value of 0.003021 for the framework 5. The median of the difference between the population proportion classified by age of the 2017-2019 R-census and T-census were 0.71% and 0.54% respectively.

**Table 8.** Chi-square Test for Homogeneity with respect to age comparing between the 2019-2019 R-census and the T-census and Chi-square histogram distance

| Framework | Age | R-census frequency (percentage) | T-census frequency (percentage) | Absolute Percentage difference | Chi-Square Statistics* (df) | P-value | CHD (AGE) |
|---|---|---|---|---|---|---|---|
| 1-4 | 0-4 | 10,476 (2.50%) | 3,274 (1.79%) | 0.71% | | | |
| | 5-9 | 21,679 (5.17%) | 8,026 (4.40%) | 0.77% | | | |
| | 10-14 | 24,003 (5.72%) | 8,891 (4.87%) | 0.85% | | | |
| | 15-19 | 24,022 (5.73%) | 9,400 (5.15%) | 0.58% | | | |
| | 20-24 | 27,083 (6.46%) | 9,564 (5.24%) | 1.22% | | | |
| | 25-29 | 28,421 (6.78%) | 11,511 (6.30%) | 0.48% | | | |
| | 30-34 | 29,016 (6.92%) | 11,830 (6.48%) | 0.44% | | | |
| | 35-39 | 32,543 (7.76%) | 13,652 (7.48%) | 0.28% | 2,149.47 (14) | < 0.001 | 0.004243 |
| | 40-44 | 34,574 (8.24%) | 15,026 (8.23%) | 0.01% | | | |
| | 45-49 | 35,294 (8.41%) | 15,348 (8.40%) | 0.01% | | | |
| | 50-54 | 36,607 (8.73%) | 16,848 (9.23%) | 0.50% | | | |
| | 55-59 | 31,547 (7.52%) | 16,095 (8.81%) | 1.29% | | | |
| | 60-64 | 25,778 (6.15%) | 12,962 (7.10%) | 0.95% | | | |
| | 65-69 | 19,939 (4.75%) | 10,461 (5.73%) | 0.98% | | | |
| | >= 70 | 38,510 (9.18%) | 19,726 (10.80%) | 1.62% | | | |
| 5 | 0-4 | 8,402 (2.14%) | 3,274 (1.79%) | 0.35% | | | |
| | 5-9 | 19,727 (5.02%) | 8,026 (4.40%) | 0.62% | | | |
| | 10-14 | 22,206 (5.65%) | 8,891 (4.87%) | 0.78% | | | |
| | 15-19 | 22,381 (5.69%) | 9,400 (5.15%) | 0.54% | | | |
| | 20-24 | 25,028 (6.37%) | 9,564 (5.24%) | 1.13% | | | |
| | 25-29 | 26,271 (6.68%) | 11,511 (6.30%) | 0.38% | | | |
| | 30-34 | 26,887 (6.84%) | 11,830 (6.48%) | 0.36% | | | |
| | 35-39 | 30,431 (7.74%) | 13,652 (7.48%) | 0.26% | 1,507.43 (14) | < 0.001 | 0.003021 |
| | 40-44 | 32,696 (8.32%) | 15,026 (8.23%) | 0.09% | | | |
| | 45-49 | 33,402 (8.5%) | 15,348 (8.40%) | 0.10% | | | |
| | 50-54 | 34,879 (8.87%) | 16,848 (9.23%) | 0.36% | | | |
| | 55-59 | 30,158 (7.67%) | 16,095 (8.81%) | 1.14% | | | |
| | 60-64 | 24,719 (6.29%) | 12,962 (7.10%) | 0.81% | | | |
| | 65-69 | 19,189 (4.88%) | 10,461 (5.73%) | 0.85% | | | |
| | >= 70 | 36,658 (9.33%) | 19,726 (10.80%) | 1.47% | | | |

* Chi-Square Test for Homogeneity

**3.3 Comparing the 5 conceptual frameworks on the use of the single database (2019 R-census) and the multiple databases (2017-2019 R-census)**

The above results are summarized and compared by ranking the measures in which the 1st Rank represents a framework with the best measure and the highest rankings indicates a framework with the worst measure. Thus, the lowest average rank across the measures will indicate the best R-census. The frameworks based on the registration databases (the 2019 R-census and the 2017-2019 R-census) are divided into 3 groups: group 1 includes the framework 1-4 from the 2019 R-census, group 2 includes the framework 1-4 from the 2017-2019 R-census, and group 3 includes the framework 5 from the 2017-2019 R-census. Four measures were ranked including coverage rate, overcoverage rate, Chi-Square Histogram Distance (sex), and Chi-Square Histogram Distance (age). The results of ranking the measures are shown in Table 9.

**Table 9.** Coverage Rate, Overcoverage Rate, Chi-Square Histogram Distance (sex), and Chi-Square Histogram Distance (age) and ranks

|  | Coverage rate | Overcoverage rate | CHD(Sex) | CHD(Age) | Average rank |
| --- | --- | --- | --- | --- | --- |
| Framework 1-4 from the 2019 R-census | 0.688496 (2) | 0.655908 (1) | 0.00006257 (1) | 0.003051 (2) | **1.5** |
| Framework 1-4 from the 2017-2019 R-census | 0.720602 (1) | 0.686306 (3) | 0.00006970 (3) | 0.004243 (3) | 2.5 |
| Framework 5 from the 2017-2019 R-census | 0.681585 (3) | 0.683317 (2) | 0.00006784 (2) | 0.003021 (1) | 2.0 |

Note. The number in the parathesis is the rank where 1 indicate the best R-census based on a particular measure.

From Table 9, the smallest average rank (1.5) of measures was in framework 1-4 from the 2019 R-census. Thus, the conceptual framework 1-4 from the 2019 R-census was the framework that provided the most comprehensive measure that aligns with the 2019 traditional census from the National Statistical Office. This framework provides the lowest overcoverage rate which reflects the smallest amount of people outside the coverage area (Chachoengsao province) participating in the census. However, the overcoverage rate of 0.656 or 65.6% is still considered high. Additionally, the conceptual framework 1-4 in the 2019 R-census yields the lowest CHD (sex) value, indicating the smallest difference in the population proportion classified by sex between the 2019 R-census and the 2019 traditional census, with a median of the difference in sex proportion of only 0.79%.

**4. Conclusion and discussion**

Comparing the framework 1 to framework 5 on using a single and multiple databases, we found the identical results between framework 1-4 which is a result from the completion of data with respect to the main census variables including PID, First name, Last name, year of birth, prefix of name, and sex. Comparing between framework 1-4 and framework 5 on using a single and multiple databases, using framework 1-4 results in a better coverage rate and better data distribution regarding to sex. In this study, we examine a quality of the framework as individual framework. As mentioned earlier, many countries use several frameworks to conduct the R-census. Future research may incorporate many frameworks and test the quality of those combined framework.

Regarding to the number of databases, it is surprisingly that using only one recent database provide better data quality (lower overcoverage rate and better data distribution regarding to sex) than using multiple databases. This might be because the data especially the main census variables is quite complete, i.e. no item missing data. Moreover, many people might migrate to other areas outside the census area before the census date. This might cause the use of multiple databases from previous years to have higher overcoverage. This study used only Basic Minimum Need (BMN) database as the register based. Future research may consider using more than one source of database as well as to use databases that are reliable to a specific variable. For example, variables related to health should come from a database provided by the Ministry of Public Health. However, we must certain that those databases contain main fields used to merge every database. Moreover, variables used in each source must have the identical meaning as the census variable. For example, the word "household" has been defined differently across government agencies in Thailand. As such, we must be careful to use the variable as well-defined as census variable.

Resources including people and infrastructure as well as regulation should be prepared to support the register-based census. Conducting register-based census requires people who have database and computer skills to do the data preparation and integration. Infrastructure including multiple servers for distributed storages and processing of big data should be well prepared to handle data streaming from multiple sources. Moreover, required by laws, all sensitive data must be protected and can be accessed by only legitimate parties. Hence, infrastructure must have high-security access control and data protection mechanism. Having these resources ready will facilitate the country to do a register-based census.

A regulation that prevents using database from each governmental agency should be modified so that it can facilitate data sharing to conduct the register-based census. In addition, the government's data security and privacy policies are the crucial factors in ensuring public trust. Many register-based censuses use administrative data that are collect for specific purposes, such as healthcare, and may not inform a purpose of census enumeration to the people whom they are collecting data from. The government agency should definitely inform the purpose of using their data in the register-based census and ensure their data privacy and data security.